\documentclass[reprint, superscriptaddress, amsmath,amssymb,
 aps, prx]{revtex4-2}

\usepackage{siunitx}
\usepackage{graphicx}% Include figure files
\usepackage{dcolumn}% Align table columns on decimal point
\usepackage{bm}% bold math
\usepackage{braket}
\usepackage{comment}
\usepackage{upgreek} % for units
\usepackage{subfiles}
\usepackage{hyperref}
\usepackage{cleveref}

\usepackage{graphicx}% Include figure files
\usepackage{dcolumn}% Align table columns on the decimal point
\usepackage{bm}% bold math
\usepackage{lipsum}
\usepackage{braket}
\usepackage{siunitx}
\usepackage{epigraph} 
\usepackage{algorithm}
\usepackage{algorithm}% http://ctan.org/pkg/algorithms
\usepackage{algpseudocode}% http://ctan.org/pkg/algorithmicx

\usepackage{tikz}
\usepackage{xcolor}

\DeclareRobustCommand{\stopwatchicon}{%
  \tikz[baseline=-0.6ex, line cap=round, line join=round, x=1em, y=1em]{
    \def\s{0.95} % overall size
    \draw[line width=0.09em] (0,0) circle[radius=0.42*\s];
    \draw[line width=0.09em] (-0.10*\s,0.46*\s) -- (0.10*\s,0.46*\s);
    \draw[line width=0.09em] (0.00*\s,0.46*\s) -- (0.00*\s,0.56*\s);
    \draw[line width=0.09em] (0.30*\s,0.38*\s) -- (0.40*\s,0.48*\s);
    \draw[line width=0.09em] (0,0) -- (0,0.22*\s);
    \draw[line width=0.09em] (0,0) -- (0.18*\s,-0.10*\s);
  }%
}

\newcommand{\Tstop}{T_{\!\text{\stopwatchicon}}}

\newcommand{\ketone}{{\ensuremath{|1\rangle}}}

\begin{document}
\preprint{APS/123-QED}

%%%%%%%%%%%%%%%%%%%%%
%%   DOC SETUP     %%
%%%%%%%%%%%%%%%%%%%%%
\title{Millisecond-Scale Calibration and Benchmarking of Superconducting Qubits}

\author{Malthe A. Marciniak}
\email{malthe.asmus.nielsen@nbi.ku.dk}
\affiliation{Center for Quantum Devices, Niels Bohr Institute, University of Copenhagen, Denmark}
\affiliation{NNF Quantum Computing Programme, Niels Bohr Institute, University of Copenhagen, Denmark}

\author{Rune T. Birke}
\affiliation{Center for Quantum Devices, Niels Bohr Institute, University of Copenhagen, Denmark}
\affiliation{NNF Quantum Computing Programme, Niels Bohr Institute, University of Copenhagen, Denmark}
\affiliation{Center for the Mathematics of Quantum Theory, MATH Department, University of Copenhagen}

\author{Johann B. Severin}
\affiliation{Center for Quantum Devices, Niels Bohr Institute, University of Copenhagen, Denmark}
\affiliation{NNF Quantum Computing Programme, Niels Bohr Institute, University of Copenhagen, Denmark}

\author{Fabrizio Berritta}
\affiliation{Center for Quantum Devices, Niels Bohr Institute, University of Copenhagen, Denmark}
\affiliation{NNF Quantum Computing Programme, Niels Bohr Institute, University of Copenhagen, Denmark}
\affiliation{Research Laboratory of Electronics, Massachusetts Institute of Technology, Cambridge, MA 02139, USA}
\affiliation{Institute of Experimental and Applied Physics, University of Regensburg, 93040 Regensburg, Germany}

\author{Daniel Kjær}
\affiliation{NNF Quantum Computing Programme, Niels Bohr Institute, University of Copenhagen, Denmark}

\author{Filip Nilsson}
\affiliation{NNF Quantum Computing Programme, Niels Bohr Institute, University of Copenhagen, Denmark}

\author{Smitha N. Themadath}
\affiliation{NNF Quantum Computing Programme, Niels Bohr Institute, University of Copenhagen, Denmark}

\author{Sangeeth Kallatt}
\affiliation{NNF Quantum Computing Programme, Niels Bohr Institute, University of Copenhagen, Denmark}

\author{James L. Webb}
\affiliation{NNF Quantum Computing Programme, Niels Bohr Institute, University of Copenhagen, Denmark}

\author{Kristoffer Bentsen}
\affiliation{NNF Quantum Computing Programme, Niels Bohr Institute, University of Copenhagen, Denmark}

\author{Tonny Madsen}
\affiliation{Center for Quantum Devices, Niels Bohr Institute, University of Copenhagen, Denmark}
\affiliation{NNF Quantum Computing Programme, Niels Bohr Institute, University of Copenhagen, Denmark}

\author{Zhenhai Sun}
\affiliation{Center for Quantum Devices, Niels Bohr Institute, University of Copenhagen, Denmark}
\affiliation{NNF Quantum Computing Programme, Niels Bohr Institute, University of Copenhagen, Denmark}

\author{Svend Krøjer}
\affiliation{Center for Quantum Devices, Niels Bohr Institute, University of Copenhagen, Denmark}
\affiliation{NNF Quantum Computing Programme, Niels Bohr Institute, University of Copenhagen, Denmark}

\author{Christopher W. Warren}
\affiliation{Center for Quantum Devices, Niels Bohr Institute, University of Copenhagen, Denmark}
\affiliation{NNF Quantum Computing Programme, Niels Bohr Institute, University of Copenhagen, Denmark}

\author{Jacob Hastrup}
\affiliation{Center for Quantum Devices, Niels Bohr Institute, University of Copenhagen, Denmark}
\affiliation{NNF Quantum Computing Programme, Niels Bohr Institute, University of Copenhagen, Denmark}

\author{Morten Kjaergaard}
\email{mkjaergaard@nbi.ku.dk}
\affiliation{Center for Quantum Devices, Niels Bohr Institute, University of Copenhagen, Denmark}
\affiliation{NNF Quantum Computing Programme, Niels Bohr Institute, University of Copenhagen, Denmark}

\date{\today}

%%%%%%%%%%%%%%%%%%%%%
%%    ABSTRACT     %%
%%%%%%%%%%%%%%%%%%%%%
\begin{abstract}
Superconducting qubit parameters drift on sub-second timescales, motivating calibration and benchmarking techniques that can be executed on millisecond timescales.
We demonstrate an on-FPGA workflow that co-locates pulse generation, data acquisition, analysis, and feed-forward, eliminating CPU round trips.
Within this workflow, we introduce sparse-sampling and on-FPGA inference tools, including computationally efficient methods for estimation of exponential and sine-like response functions, as well as on-FPGA implementations of Nelder–Mead optimization and golden-section search.
These methods enable low-latency primitives for readout calibration, spectroscopy, pulse-amplitude calibration, coherence estimation, and benchmarking.
We deploy this toolset to estimate $T_1$ in 10 ms, optimize readout parameters in 100 ms, optimize pulse amplitudes in 1 ms, and perform Clifford randomized gate benchmarking in 107 ms on a flux-tunable superconducting transmon qubit.
Running a closed-loop on-FPGA recalibration protocol continuously for 6 hours enables more than 74,000 consecutive recalibrations and yields gate errors that consistently retain better performance than the baseline initial calibration.
Correlation analysis shows that recalibration suppresses coupling of gate error to control-parameter drift while preserving a coherence-linked performance.
Finally, we quantify uncertainty versus time-to-decision under our sparse sampling approaches and identify optimal parameter regimes for efficient estimation of qubit and pulse parameters.
\end{abstract}

%%%%%%%%%%%%%%%%%%%%%
%%      TITLE      %%
%%%%%%%%%%%%%%%%%%%%%
\maketitle

%%%%%%%%%%%%%%%%%%%%%
%%  INTRODUCTION   %%
%%%%%%%%%%%%%%%%%%%%%
High-fidelity quantum control relies on accurate calibration of both qubit gate and readout parameters~\cite{Krantz2019AQubits, Arute2019QuantumProcessor, Wittler2021IntegratedQubits, Werninghaus2021High-SpeedReset, Tornow2022MinimumRates, Walter2017RapidQubits, Fan2025CalibratingProcessor}. 
However, calibration itself is challenged by random fluctuations~\cite{Klimov2018FluctuationsQubits, Carroll2022DynamicsTimes, Burnett2019DecoherenceQubits}, drifts of the environment~\cite{proctor2020DetectingTrackingDrift}, and high-energy events introducing correlated errors~\cite{mcewen2022cosmicrays, liCosmicrayinducedCorrelatedErrors2025} affecting the qubit control parameters. 
Each of these effects on the environment occur over different timescales from hours in the case of drifts, to tens of seconds in the case of high-energy events.
Recently, it has been observed that fluctuations of the environment in superconducting qubits can even occur on millisecond timescales~\cite{Berritta2025Real-timeQubits}. 
This raises the question of what it means for a qubit to be calibrated. 
In many laboratory workflows, calibration and validation routines can take minutes to hours~\cite{Werninghaus2021High-SpeedReset, Tornow2022MinimumRates}, as parameters are estimated from long acquisitions to reduce statistical uncertainty. 
Device parameters extracted on timescales comparable to these fluctuations can only represent average performance over this non-static environment. 
Robust pulse design can mitigate sensitivity to drift, but often introduces trade-offs in control performance~\cite{Wright2025SuperconductingFluctuations, Kuzmanovic2024High-fidelityPulses, Werninghaus2021High-SpeedReset}.
This motivates calibration loops that are fast enough to track the relevant drifts while maintaining high-fidelity operation~\cite{Berritta2025EfficientTracking, Berritta2024Physics-informedFluctuations}.

Several techniques have been used to speed up the repetition rate of calibration putting experiments closer to the timescales of the environment. Significant speed-ups can be achieved with efficient qubit reset schemes ~\cite{risteFeedbackControlSolidState2012,salatheLowLatencyDigitalSignal2018, Geerlings2013DemonstratingQubit, magnardFastUnconditionalAllMicrowave2018,eggerPulsedResetProtocol2018,miaoOvercomingLeakageQuantum2023,Rol2017RestlessGates,Werninghaus2021High-SpeedReset,Tornow2022MinimumRates, reuerRealizingDeepReinforcement2023}, eliminating down-time from waiting for qubits to naturally relax. The next bottleneck for calibration speed then becomes offloading of the data to a central measurement computer, analysis of the results and updating of relevant parameters.
Low-latency workflows embedded onto Field-Programmable Gate Arrays (FPGAs), facilitated by recent advances in commercial pulse control hardware, are becoming increasingly common for efficient learning and estimation of qubit parameters~\cite{gebhartLearningQuantumSystems2023,Berritta2025EfficientTracking, Berritta2024Physics-informedFluctuations, Berritta2025Real-timeQubits}. These controllers can estimate key parameters on hardware in real time, avoiding host-computer latency in the data acquisition cycle and allowing for active feedback to stabilize qubit operations in the presence of the fluctuating environment~\cite{dumoulinstuyckSiliconSpinQubit2024,parkPassiveActiveSuppression2025}.
Additionally, gains in calibration speed can be achieved by efficient selection of experiment sampling points, optimizing the amount of information gained per experimental shot.

In this work, we demonstrate an on-FPGA calibration and analysis workflow that reduces the time-to-decision of key single-qubit calibration primitives to the tens-of-milliseconds regime implemented on a Quantum Machines OPX1000 control system. 
Here, \textit{on-FPGA} means that pulse generation, measurements, parameter estimation, optimization, fitting, and feed-forward are all executed on the FPGA control hardware in one loop.
In the conventional \textit{offloading} approach, measurement outcomes are transferred to a central computer for analysis before updated parameters are returned to the controller. 
By avoiding this round trip, the on-FPGA workflow removes the dominant communication and data-loading overhead associated with standard analysis workflows.

Using this control framework, we make three contributions. 
First, in Section~\ref{sec:drift}, we highlight how rapidly the qubit environment drifts and show that the conventional offloading workflow is fundamentally mismatched to these sub-second dynamics, since even LAN round-trip latency is typically on the order of tens of milliseconds and must be incurred for each update, making it incompatible with maintaining near-optimal calibration in real time.
Second, in Section~\ref{sec:toolkit}, we present an on-FPGA library of calibration primitives that execute end-to-end on control hardware with millisecond time-to-decision.
This includes two novel sparse-sampling methods for SPAM-independent extraction of exponential and sinusoidal responses using closed-form analytical expressions enabling memory-light analyses suitable for FPGA implementation. 
We deploy these methods for efficient extraction of $T_1$, CRB fidelity, pulse-train-based amplitude optimization and Ramsey-based frequency estimation. 
Additionally, we demonstrate FPGA implementations of general-purpose $N$-dimensional Nelder-Mead optimization algorithm \cite{Nelder1965AMinimization} which we apply for readout optimization, as well as one-dimensional optimization via golden-section search \cite{Kiefer1953SequentialMaximum, Avriel1966OptimalityTechnique}, suitable for, e.g., peak-finding in spectroscopy experiments. 
Third, in Section~\ref{sec:labtime}, we demonstrate stable long-term operation enabled by this framework. 
Finally, we  profile the timing budget of each primitive in Section~\ref{sec:timing} and quantify the trade-off between time-to-decision and estimator precision under sparse sampling in Section~\ref{sec:uncertainty}.

%%%%%%%%%%%%%%%%%%%%%
%%    FIGURE 1     %%
%%%%%%%%%%%%%%%%%%%%%
\section{Sub-second coherence dynamics}
\label{sec:drift}

\begin{figure}[t]
    \centering
    \includegraphics[width=\linewidth]{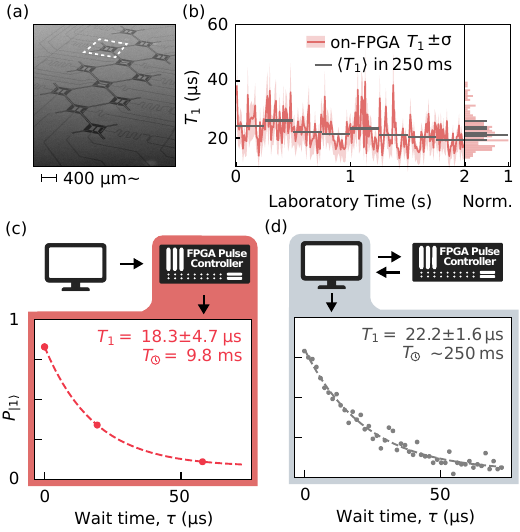}
    \caption{
    Rapid parameter estimation on a superconducting quantum processor. 
    (a) Tilted SEM micrograph of the 12-qubit flux-tunable transmon qubit processor. The specific qubit used in this work is highlighted in a white dashed box. 
    (b) Continuous $T_1$ tracking using our on-FPGA control, with roughly 200 $T_1$ estimates performed in 2~seconds of wall-clock time.
    (c) A $T_1$ measurement (red points) measured with our on-FPGA loop performing both measurement and estimation within a closed-loop taking 9.8 ms of wallclock time.
    (d) A $T_1$ measurement of the qubit performed with a conventional offloading loop, taking roughly $\sim 250\,\mathrm{ms}$ of wall-clock time, and dominated by data transfer, dense sampling and software overhead.}
    \label{fig:1}
\end{figure}

Here we motivate millisecond-scale calibration and benchmarking by examining fluctuations in the coherence of a superconducting qubit. 
Fig.~\ref{fig:1}(a) shows an SEM image of the superconducting device used in this work. 
The device comprises 12 flux-tunable transmon qubits with fixed qubit--qubit coupling, each capacitively coupled to a dedicated resonator for dispersive readout.
The qubit highlighted by the white square is used throughout this manuscript, and all reported data are obtained from experiments on that qubit. 
Further details on the device and the cryogenic setup are presented in Appendix \ref{app:experimental-setup}.

In Fig.~\ref{fig:1}(b), we show $T_1$ measurements fluctuating over a two second acquisition window. 
These rapid fluctuations are representative of other recent results on sub-second dynamics of the environment of superconducting qubits~\cite{Berritta2025Real-timeQubits}. 
The fast fluctuations are obtained by the on-FPGA sparse estimation method shown in Fig.~\ref{fig:1}(c) (this method is presented in more detail in Section~\ref{sec:ADE}). 
In this protocol, each $T_1$ is obtained by uploading a three-point sparse-sampling decay estimator function alongside the pulse sequence to the FPGA controller.
The FPGA acquires and averages the resulting measurements and then evaluates the $T_1$ estimation function, yielding in this case $T_1=18.3\pm4.7\,\mu\mathrm{s}$. 
This estimate is obtained in $\Tstop = 9.8\,\mathrm{ms}$, where we define the time-to-decision, $\Tstop$, as the end-to-end time from launching the first experimental shot to the moment a parameter estimate of $T_1$ is available to the controller. This includes sequence execution, data transfer, and analysis.

To benchmark our on-FPGA approach we measure the $T_1$ also using the conventional offloading of data with more densely sampled points, to a central CPU that performs the fitting, shown in Fig.~\ref{fig:1}(d). 
The estimated $T_1$ is $22.2\pm1.6\,\mu\mathrm{s}$ with a time-to-decision $\Tstop \sim 250\,\mathrm{ms}$. 
This measured $T_1$ is representative of the average qubit performance as can be seen in the fluctuations of Fig.~\ref{fig:1}(b). 
The grey lines represent 250\,ms acquisition windows of the conventional $T_1$.
Each estimate using the conventional approach would correspond to the eight estimates (grey lines) of the fluctuations in Fig.~\ref{fig:1}(b). 
When working in the conventional data-offloading paradigm the acquisition latency is largely set by data-transfer and data-loading overheads rather than by the pulse sequence and repetition rate of the estimation protocol and sets a limit on how fast a calibration protocol can be run. 
This limit is much slower than the typical fluctuations that the qubit's environment experiences.
This motivates shifting from a calibration loop that involve multiple round-trips between the controller and a separate analysis/feed-forward CPU to a fully on-FPGA enabled workflow.
In the following sections we show how co-locating pulse generation, acquisition, analysis, and feed-forward on the FPGA enables autonomous single-qubit recalibration with millisecond time-to-decision.

%%%%%%%%%%%%%%%%%%%%%
%%    FIGURE 2     %%
%%%%%%%%%%%%%%%%%%%%%
\begin{figure*}[t]
    \centering
    \includegraphics[width=\linewidth]{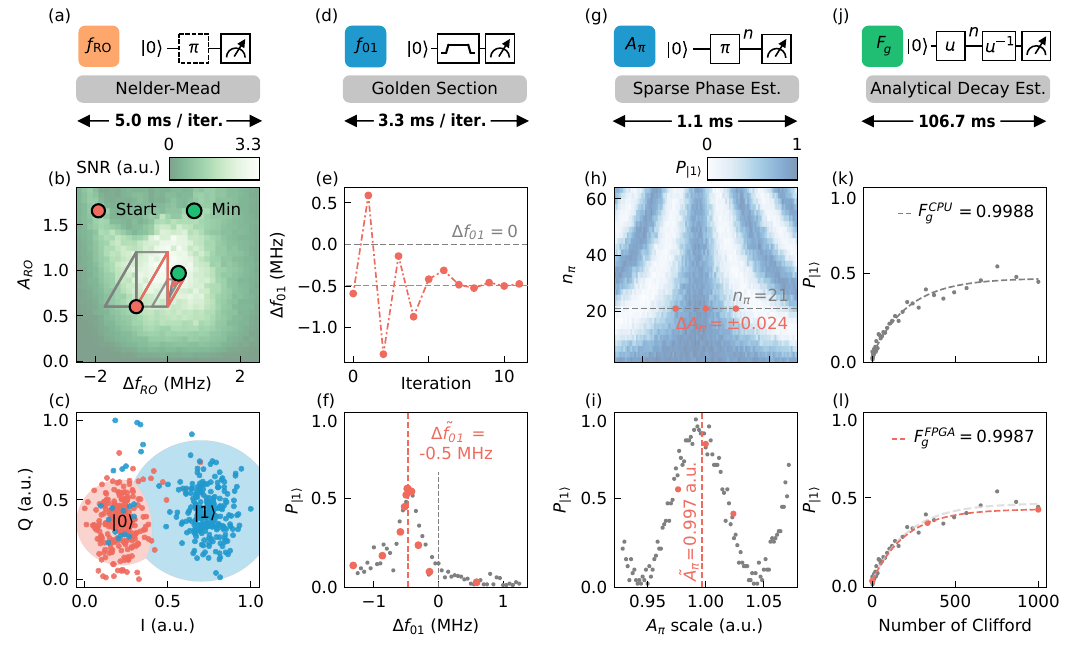}
\caption{Offloading versus on-FPGA calibration primitives. 
(a--c) Readout optimization. 
(a) Readout-optimization primitive and time-to-decision per iteration. 
(b) SNR landscape in $(\Delta f_{\mathrm{RO}},A_{\mathrm{RO}})$ with a representative Nelder--Mead trajectory (red) over a conventional sweep (green heatmap). Gray lines show the Nealder--Mead simplexes.
(c) Example single-shot IQ clusters used to evaluate the SNR objective. 
(d--f) Drive-frequency refinement by spectroscopy. 
(d) Spectroscopy primitive and per-iteration time-to-decision. 
(e) Golden-section-search iteration trace to maximize $P_{\ket{1}}$ over $\Delta f_{01}$.
(f) Spectroscopy data comparing a dense grid sampled with offloading strategy (grey) to golden-section search executed on-FPGA (red). 
(g--i) $\pi$-pulse amplitude correction. 
(g) $\pi$-train primitive and time-to-decision. 
(h) Conventional amplitude scan (blue heatmap) with the three-point estimator sampling points highlighted. 
(i) Three-point amplitude update using $n_\pi=21$ (red) over a line cut of (h) (grey).
(j--l) Clifford randomized benchmarking. 
(j) CRB primitive and time-to-decision. 
(k) Example traditional offloading-workflow CRB analysis (dense sampling and fit functions executed offline).
(l) On-FPGA CRB analysis using analytical decay estimation from three sequence lengths (red) overlaid on CRB data from offloading workflow (grey), see text for details. 
}
    \label{fig:2}
\end{figure*}

%%%%%%%%%%%%%%%%
% FPGA TOOLKIT %
%%%%%%%%%%%%%%%%
\section{Rapid FPGA-based Calibration, Analysis and Optimization Toolkit}
\label{sec:toolkit}

\subsection{Scope and design principles}
To operate on millisecond timescales, we implement a set of calibration primitives that execute end-to-end on the FPGA, as summarized in Fig.~\ref{fig:2}. 
We assume the device is \emph{approximately} calibrated at the outset, meaning that a usable readout and coarse $\pi$ and $\pi/2$ pulses are available.
This allows approximate preparation of $\lvert 0\rangle$ and $\lvert 1\rangle$ and provides the starting point for recalibration.
Throughout this section, we characterize the speed of a calibration primitive by its time-to-decision, $\Tstop$.
Each primitive is designed to minimize $\Tstop$ by combining sparse measurements with on-FPGA processing, analysis and feed-forward.

%%%%%%%%%%
% T1/ADE %
%%%%%%%%%%
\subsection{Qubit coherence tracking with analytical decay estimation}\label{sec:ADE}
Many characterization experiments, including $T_1$ and randomized benchmarking, result in exponential decay curves for which the decay rate is the parameter of interest.
Therefore, to enable rapid tracking of decay rates with minimal acquisition overhead, we introduce \emph{Analytical Decay Estimation} (ADE), a method that extracts an exponential decay rate from only three sampling points. 
ADE does not rely on fitting or minimization and is therefore naturally suited to fast and memory-efficient on-FPGA execution. 
A detailed exposition of ADE is provided in Appendix \ref{app:sparse-exp}. 
Here exemplify ADE for the case of $T_1$ estimation, while in Sec.~\ref{sec:crb} we reuse the same method to infer gate fidelity parameters in randomized benchmarking.

For ADE $T_1$ tracking, we measure the excited-state probability $P_\ketone(\tau)$ at three delays $\tau\in\{t_0,\,t_0+\Delta t,\,t_0+3\Delta t\}$ and compute the decay rate $\Gamma_1=1/T_1$ from a closed-form expression:
\begin{equation}
\Gamma_1 = -\frac{1}{\Delta t}\ln\!\left(\sqrt{c(t_0,\Delta t) - \tfrac{3}{4}} - \tfrac{1}{2}\right), 
\label{eq:ADE}
\end{equation}
where 
\begin{equation}
c(t_0,\Delta t) = \frac{P_\ketone(t_0 + 3\Delta t) - P_\ketone(t_0)}{P_\ketone(t_0 + \Delta t) - P_\ketone(t_0)}.\label{eq:k_ADE}
\end{equation}
Uncertainty can be obtained by bootstrapping from measurement shots or by standard error propagation through Eq.~\eqref{eq:ADE}, with standard deviations of the three probability measurements obtained from measurement shot-noise. 

Note that since ADE extracts the decay rate independently of the offset and amplitude of the exponential, the method is SPAM-independent.

In the implementation used for Fig.~\ref{fig:1}(b), we choose delays $\tau\in\{16\,\mathrm{ns},\,16\,\mathrm{ns}+\widetilde{T}_1,\,16\,\mathrm{ns}+3\widetilde{T}_1\}$, where $\widetilde{T}_1$ denotes the most recently estimated relaxation time, and sample $50$ shots at each delay. 
Using these parameters, we obtain a representative relaxation time of $T_1 = 18.3\pm 4.7\,\mu\mathrm{s}$ with a total time-to-decision of $\Tstop = 9.8$ms.

%%%%%%%%%%%%%%%
% Nelder-Mead %
%%%%%%%%%%%%%%%
\subsection{Readout optimization via two-dimensional minimizer}
Readout quality is essential for high quantum processor performance.
In Fig.~\ref{fig:2}(a) we illustrate the pulse-level primitive for dispersive readout optimization: A readout pulse is played after initializing the qubit in either the ground or excited state.
The goal is to maximize state distinguishability by finding optimal readout-pulse detuning $\Delta f_{\mathrm{RO}}$ and amplitude $A_{\mathrm{RO}}$. 
At every candidate point, $(\Delta f_{\mathrm{RO}},A_{\mathrm{RO}})$, we acquire batches of single-shot, integrated IQ samples for state preparation in $\lvert 0\rangle$ and $\lvert 1\rangle$ and compute a scalar readout-quality objective from the resulting IQ clusters. 
Here, we use the signal-to-noise ratio (SNR) metric defined in Appendix~\ref{app:readout-snr}~\cite{Walter2017RapidQubits,Heinsoo2018RapidQubits,Mallet2009Single-shotElectrodynamics} as our objective. 
An example of the state-separated IQ clusters used for the SNR evaluation is shown in Fig.~\ref{fig:2}(c).
At this stage of the calibration protocol we intentionally do not rely on active reset, since readout optimization is typically performed without a trusted IQ-discrimination threshold (and, by extension, reset conditioned on measured state), making the procedure more representative but leading to increased per-iteration times.

The heatmap of Fig.~\ref{fig:2}(b) shows a conventional two-dimensional sweep in the $\text{offloading}$ workflow, where the SNR is evaluated on a dense grid of $(\Delta f_{\mathrm{RO}},A_{\mathrm{RO}})$ points, requiring a total runtime of $\Tstop \approx 6\,\mathrm{s}$ to complete.

To find the optimal SNR point in $(\Delta f_{\mathrm{RO}},A_{\mathrm{RO}})$--space, we implement a Nelder--Mead optimizer~\cite{Nelder1965AMinimization} within the on-FPGA workflow. 
Rather than evaluating a dense grid, the optimizer proposes new $(\Delta f_{\mathrm{RO}},A_{\mathrm{RO}})$ points, triggers the corresponding measurements, computes the SNR, and updates the simplex without offloading the data for analysis on a separate computer. 
A representative optimization trajectory found with this approach is shown in red in Fig.~\ref{fig:2}(b) (gray lines indicate the simplexes proposed by the Nelder--Mead optimizer). 
In this implementation, each Nelder--Mead iteration (one objective evaluation and simplex update) takes $\sim 5.0\,\mathrm{ms}$. For the example in Fig.~\ref{fig:2}(b), the optimizer converges in $20$ iterations, yielding optimized readout parameters in $\Tstop=100$ ms, i.e., with substantially reduced end-to-end runtime compared to the full two-dimensional sweep. 
The stopping criteria used by the algorithm is summarized in Appendix~\ref{app:stopping}.

%%%%%%%%%%%%%%%%%%
% Golden section %
%%%%%%%%%%%%%%%%%%
\subsection{Spectroscopic peak-finding via golden-section search}
Finding peaks in spectroscopic experiments is a common task in superconducting qubit characterization.
In Fig.~\ref{fig:2}(d) we show a typical qubit spectroscopy pulse sequence.
A saturating pulse is applied on the qubit to put it in a mixed state, followed by a projective readout.
Fig.~\ref{fig:2}(f) shows qubit spectroscopy used to determine the qubit transition frequency by maximizing the measured excitation probability, $P_{\ketone}$, as a function of detuning $\Delta f_d = \widetilde f_q - f_d$ (grey points), where $\widetilde f_q$ is the most recent available estimate for the qubit frequency, and $f_d$ is the drive frequency. 
In the offloading workflow, the resonance is typically located by evaluating the response on a fixed, pre-defined frequency grid, followed by a peak-finder or reasonable fit function, which requires sampling at many frequency points thus incurring a larger time-to-decision.

To reduce this overhead, we implement golden-section search (GSS) within the on-FPGA workflow~\cite{Kiefer1953SequentialMaximum,Avriel1966OptimalityTechnique}. 
GSS has a deterministic control flow and requires minimal memory, making it well suited to our on-FPGA scheme. Further details on GSS are given in Appendix \ref{app:gss}. 
In our implementation, for each iteration, we apply a $10\,\mu\mathrm{s}$ drive followed by readout at a candidate detuning, acquire multiple shots, evaluate $P_{\ketone}$, and update the bracketing interval.
The resulting sequence of frequency queries is shown by the red points in Fig.~\ref{fig:2}(f), with the corresponding GSS convergence trace shown in Fig.~\ref{fig:2}(e). 

Using this approach, the transition frequency is identified in a total time-to-decision of $\Tstop \approx 39\,\mathrm{ms}$, corresponding to $3.3\,\mathrm{ms}$ per iteration.

%%%%%%%%%%%%
% pi-pulse %
%%%%%%%%%%%%
\subsection{Pulse-amplitude correction and frequency correction via sparse phase estimation}\label{sec:3pointsampling}
Another class of experiments yield sinusoidal responses for which the parameter of interest is the phase. For example, in Fig.~\ref{fig:2}(g) we show a typical single-qubit pulse-train scheme used to amplify amplitude errors in a $\pi$-pulse (the $\pi$-train approach).
In a conventional calibration workflow, these pulse-amplitude errors are often characterized by scanning a range of $\pi$-pulse amplitudes (shown as a heatmap in Fig.~\ref{fig:2}(h)) and fitting the resulting oscillatory response.
This approach is fairly robust, but conventionally data-intensive in both measurements and evaluation, resulting in a longer time-to-decision of $\Tstop \approx 1.2\,\mathrm{s}$, in our specific case.

To reduce acquisition overhead while retaining sensitivity to small amplitude errors and robustness, we implement a technique we call \textit{Sparse Phase Estimator} (SPE), which can be deployed for all sine-like response functions.
Previous demonstrations of sparse sampling estimation for pulse calibration have used two-point estimators to infer a parameter-correction from the local slope of an underlying sinusoidal response.
However, this approach requires additional information about the baseline oscillation (e.g., offset, contrast, and phase) obtained from separate measurements or fits~\cite{Rol2017RestlessGates,Tornow2022MinimumRates,Werninghaus2021High-SpeedReset}.

By contrast, our SPE approach uses three points sampled symmetrically around a central point to cancel unknown and irrelevant quadratures of the response, yielding an analytical phase correction without the need for prior calibration.
Compared to two-point updates, it offers a larger capture range and reduced sensitivity to offsets and contrast variations in the measured probabilities.
These features make our sparse phase estimator stable under drift and well suited to on-FPGA execution without relying on additional memory or pre-calibration.
In its general form, the method estimates a phase error on a sinusoidal signal by estimating the response at the current best guess of the point with no phase error (denoted $\theta_0$) and at the maximally sensitive points $\theta_\pm = \theta_0 \pm \pi/2$. 
Doing so yields an estimate to the argument of the sinusoidal signal with the form:
\begin{equation}\label{eq:pi-correction-3PE}
\theta =
\arctan2\!\left(P_{\ketone,-} - P_{\ketone,+},\,2(P_{\ketone,0}-\overline{P}_\ketone)\right),
\end{equation}
where $P_\ketone$ is the excited-state population, $P_{\ketone,[\pm,0]}$ are measured at $\theta_{[\pm,0]}$, and $\overline{P}_\ketone=(P_{\ketone,-}+P_{\ketone,+})/2$.
Sparse phase estimation can be deployed to estimate phase-offset in any response function with sine-like behaviour, such as frequency estimation in Ramsey experiments, DRAG calibration~\cite{Motzoi2009SimpleQubits}, and $\pi$- and $\pi/2$-pulse amplitude calibration.
Further details of the applications and a derivation of SPE is given in Appendix~\ref{app:sparse-sinus}.

To exemplify, we apply SPE to $\pi$-pulse amplitude correction, running a $\pi$-train of $n_\pi$ consecutive $\pi$ pulses and probe the response at three amplitude scalings of the $\pi$-pulse, $\alpha_\pm=\alpha_0\!\left(1\pm\frac{1}{2n_\pi}\right)$ and $\alpha_0$, where $\alpha_0$ denotes the current best guess of the $\pi$-pulse amplitude scaling factor. 
Sampling at these maximally sensitive points yields a correction rule
\begin{equation}
\delta \alpha =
\frac{\theta}{n_\pi} - \pi, 
\label{eq:3-point-est:pi-delta-alpha}
\end{equation}
In Fig.~\ref{fig:2}(i) (red points) we show the implementation of sparse phase estimation executed on the FPGA.
In this case, we use $n_\pi=21$ (gray points is line-cut in panel (h)), and set $\alpha_0=1$ by normalizing amplitudes to the most recently calibrated $\pi$ pulse.
The FPGA evaluates $P_{\ketone,[\pm,0]}$ and applies the corresponding amplitude update using Eqs.~\eqref{eq:pi-correction-3PE} and \eqref{eq:3-point-est:pi-delta-alpha}. 
Using this approach, $\pi$-pulse amplitude correction completes in a total time-to-decision of $\Tstop = 1.1\,\mathrm{ms}$.

%%%%%%%
% CRB %
%%%%%%%
\subsection{Analytical decay estimation applied to Clifford randomized benchmarking}\label{sec:crb}

Fig.~\ref{fig:2}(j) shows a representative single-qubit Clifford randomized benchmarking (CRB) protocol~\cite{Magesan2011ScalableProcesses,Magesan2012CharacterizingBenchmarking}. 
In standard CRB analysis, the survival probability decays approximately exponentially with the number of Clifford gates, with a general exponential form given by,
\begin{equation}
    P_\ketone(m) = C+Ap^m,
\end{equation}
where $m$ is the number of Clifford gates, and the average Clifford gate fidelity can be extracted as $F_g = (1+p)/2$ for single-qubit Cliffords. 
In Fig.~\ref{fig:2}(k) we show a representative Clifford randomized benchmarking curve analyzed by offloading data and fitting on a separate computer, with a representative gate fidelity of $F_g^\text{CPU} = 0.9988 \pm 0.0001$, measured in $\Tstop \approx 1.3\,\mathrm{s}$.

However, the exponential structure of the CRB decay makes it compatible with our closed-form three-point estimator, ADE, used for $T_1$ estimation in Section~\ref{sec:ADE}.
To deploy ADE in the context of a CRB experiment, we choose three sequence lengths $m\in\{m_0,\,m_0+\Delta m,\,m_0+3\Delta m\}$ and sample the survival probability at those points, shown in Fig.~\ref{fig:2}(l). 
The Clifford sequences are generated and randomized on-FPGA.
From these points, we estimate the decay rate via
\begin{equation}
    p = \left[ \sqrt{c(m_0,\Delta m) - \frac{3}{4}}-\frac{1}{2}\right]^{1/{\Delta m}},
\end{equation}
where
\begin{equation}
    c(m_0,\Delta m) = \frac{P_\ketone(m_0 + 3\Delta m) - P_\ketone(m_0)}{P_\ketone(m_0 + \Delta m) - P_\ketone(m_0)}.
\end{equation}
Equivalently, Eq.~(\ref{eq:k_ADE}) carries over by identifying an effective decay rate $\Gamma_p$ with the RB decay per Clifford, $\Gamma_p = -\log(p)$.

Using this on-FPGA CRB primitive with ADE, and setting $m_0= 1$ and $
\Delta m = 333$, we obtain a representative $F_g^{\mathrm{FPGA}} = 0.9987$ with a total time-to-decision of $\Tstop \approx 107\,\mathrm{ms}$.
The bulk of this time arises from the long pulse-sequences and circuit-averaging to get appropriate statistics, see Section \Ref{sec:timing}.

%%%%%%%%%
% OUTRO %
%%%%%%%%%
\subsection{Low-latency primitives as building blocks}
The primitives in Fig.~\ref{fig:2} span common calibration tasks, from readout tuning and spectroscopy to pulse-amplitude correction, coherence tracking, and benchmarking. 
In many cases, a calibration or characterization bottleneck can be recast into one of these patterns by identifying (i) a scalar objective (for minimization), or exponential-like response (ADE) or sine-like response (SPE) to estimate, and (ii) a sparse set of measurements sufficient to make a reliable update decision. 
Thus, a small set of FPGA-amenable inference and optimization motifs can cover a broad class of practical calibration problems.
In the following section we turn to applications of the on-FPGA workflow.

%%%%%%%%%%%%%%%%%%%%%
%%    FIGURE 3     %%
%%%%%%%%%%%%%%%%%%%%%
\begin{figure*}[t]
    \centering
    \includegraphics[width=\linewidth]{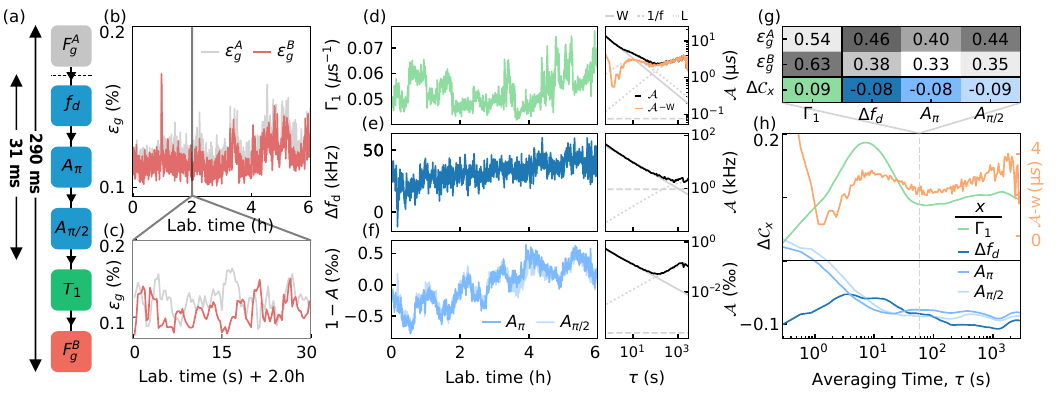}
\caption{Long-term closed-loop on-FPGA recalibration and drift-channel analysis. 
(a) Closed-loop protocol with alternating calibration and CRB validation, repeated with a $290\,\mathrm{ms}$ cadence and $\sim31\,\mathrm{ms}$ in-loop calibration latency. 
(b) Gate error over 6 hours for a static baseline, $\epsilon_g^{A}$ (grey), and continuous recalibration, $\epsilon_g^{B}$ (red), showing reduced error under closed-loop operation.
For visual clarity, a rolling average of 200 is applied to the data.
(c) 30 second zoom-in of (b), with a rolling average of 5 applied to the data.
(d--f) Concurrently tracked drift channels: $\Gamma_1=1/T_1$, inferred detuning $\Delta f_d$, and the relative $\pi$ and $\pi/2$ pulse amplitudes. 
A rolling average of 200 consecutive experiments have been applied for visual clarity.
For all three parameters, we plot Allan deviation ($\mathcal A$) in the rightmost inset, along with fitted white noise ($W$), Lorentzian ($L$) and $1/f$ noise contributions ($1/f$).
For $\Gamma_1$ we also plot the Allan deviation with white noise contribution subtracted, $\mathcal{A}-W$ (orange curve).
(g) Pearson correlations between gate error and drift channels at rolling average of 200 points (corresponding to $\tau \approx 60\,\textrm{s}$), comparing $C(\epsilon_g^{A},x)$ and $C(\epsilon_g^{B},x)$ for the four parameters tracked in this experiment.
(h) Timescale-resolved correlation difference $\Delta C(x;\tau) = C(\epsilon^B,x;\tau)-C(\epsilon^A,x;\tau)$ computed after smoothing with window $\tau$, shown alongside the Allan deviation of $\Gamma_1$ (orange, right axis), $\Delta f_\textrm{d}$ (dark blue), $A_\pi$ (blue), and $A_{\pi/2}$ (light blue). 
The dashed grey line marks the averaging time $\tau$ at which the correlations reported in panel (g) are evaluated. 
}
\label{fig:3}
\end{figure*}

\section{Continuous closed-loop on-FPGA recalibration and benchmarking}
\label{sec:labtime}

The fluctuations observed in Section~\ref{sec:drift} imply that device parameters, and hence optimal control parameters, are time-dependent on sub-second timescales. 
In particular, residual detuning, amplitude miscalibration, and drift in coherences can translate directly into variations in extracted gate fidelities. 
This motivates a feedback protocol in which calibrated parameters are refreshed at a fixed cadence and validated in situ, so that control settings track evolving device properties, even on sub-second timescales, rather than remaining static from a fixed initial point. 

Fig.~\ref{fig:3}(a) shows such a closed-loop recalibration protocol implemented fully on-FPGA. 
We begin by estimating a baseline gate fidelity $F_g^{A}$ from CRB using ADE, from a fixed, initial calibration pass. 
We then execute a calibration pass comprising pulse frequency calibration using Ramsey spectroscopy with SPE, $\pi$- and $\pi/2$-pulse amplitude calibration using pulse-trains with SPE completed within $\sim31\,\mathrm{ms}$, which we refer to as the in-loop calibration latency.
After this in-loop calibration we record a $T_1$ estimate using ADE, then repeat CRB and record a new fidelity, $F_g^B$, using the newly calibrated parameters.
This calibration/benchmarking pass is repeated with a cadence of $290\,\mathrm{ms}$, where the additional $\sim259\,\mathrm{ms}$ is set by the before/after CRB experiments and $T_1$ estimation. 
In this loop, pulse generation, data acquisition, randomization, analysis, and parameter updates are all executed on-FPGA, so that each cycle produces an updated parameter estimate without offloading data to a CPU or waiting for a fitting routine to converge before update parameters are available to the pulse controller. 

We run the closed-loop protocol continuously over a 6-hour period, performing $74{,}525$ calibration loops over this interval.
In Fig.~\ref{fig:3}(b), we plot the estimated average gate infidelity $\mathcal \epsilon_g = 1-F_g$ as a function of time over 6 hours for benchmarking with initial parameters ($\epsilon^A_g$) and continuously recalibrated parameters ($\epsilon^B_g$). 
For visual clarity the data in Fig.~\ref{fig:3}(b) shows a rolling average of 200 sequential experiments, corresponding to roughly 1 minute of wallclock time.
A zoom-in on a 30 second window is shown in Fig.~\ref{fig:3}(c) (rolling average of 5 experiments for visual clarity).
Overall, the recalibrated trace (red) remains on average lower in infidelity than the static reference (grey), indicating that frequent parameter recalibration improves performance.
Averaged over the entire timespan we achieve an average 6.4\% reduction in average gate infidelity.

To connect the observed fidelity fluctuations to underlying drift channels, we track the measured $T_1$ and control parameters. 
Fig.~\ref{fig:3}(d) shows the relaxation rate $\Gamma_1 = 1/T_1$.
Here, $T_1$ fluctuates between $14.5\,\mu\mathrm{s}$ and $27.5\,\mu\mathrm{s}$, shifting between distinct environmental states with characteristic timescales similar to previously reported $T_1$ dynamics~\cite{Klimov2018FluctuationsQubits,Carroll2022DynamicsTimes,Berritta2025Real-timeQubits}.
The average time-to-decision of the $T_1$ estimates is $\Tstop = 30$~ms.
For completeness, in Appendix~\ref{app:downsampling} we show the difference between rolling average and downsampling of these dense recalibration datasets.

Fig.~\ref{fig:3}(e) presents the drift in qubit drive frequency, reported as the inferred detuning $\Delta f_{d}$, where $\Delta f_d = 0$ corresponds to the drive frequency found from the initial calibration.
For efficient drive-frequency estimation we use SPE adapted for Ramsey experiments (see Appendix~\ref{app:sparse-sinus} for implementation details).
Over six hours we observe fluctuations of approximately $\pm25\,\mathrm{kHz}$ and with an average time-to-decision of $\Tstop = 7.5$ms. 

Finally, Fig.~\ref{fig:3}(f) shows the drift in both $\pi$- and $\pi/2$-pulse amplitudes. 
Since $A_{\pi/2} \simeq A_{\pi}/2$, estimating them independently provides a consistency check on the calibration model.
For these estimations we use SPE as presented in Section~\ref{sec:3pointsampling}.
Their close tracking over time supports reliability of the calibration sequence and validates the sparse update strategy in a closed loop.

In the right subpanels of Figs.~\ref{fig:3}(d,e,f) we calculate the Allan deviation ($\mathcal A$) of each parameter, together with fits to white noise ($W$), $1/f$ noise ($1/f$), and Lorentzian noise ($L$).
For $\Gamma_1$ we also plot the Allan deviation with its white noise contribution subtracted (orange curve, $\mathcal A - W$), which highlights the presence of a pronounced peak around $\tau \approx 10\,\textrm{s}$.
In addition, the Allan deviation of the $\pi$ and $\pi/2$ amplitude parameters also show a strong non-monotonic feature around $\tau = 10^3\,\textrm{s}$, indicating correlations across that timescale in our experimental setups.

We quantify how the gate error in the baseline and continuously recalibrated cases correlates with drifting device parameters.
Figure~\ref{fig:3}(g) reports Pearson correlations $C(\epsilon_g^i,x)$ for $x\in\{\Gamma_1,\Delta f_d,A_\pi,A_{\pi/2}\}$ and $i\in\{A,B\}$, computed from the time series in Fig.~\ref{fig:3}(b,d,e,f). 
We find $C(\epsilon_g^B,\Gamma_1) > C(\epsilon_g^A,\Gamma_1)$, indicating that under continuous recalibration the residual gate error is more tightly linked to the instantaneous coherence. 
Conversely, for $x\in\{\Delta f_d,A_\pi,A_{\pi/2}\}$ we observe $C(\epsilon_g^B,x) < C(\epsilon_g^A,x)$, consistent with recalibration continually updating these control parameters and thereby suppressing their impact on gate performance. 

To resolve the relevant timescales of correlations, in Fig.~\ref{fig:3}(h) we plot the correlation difference $\Delta C(x;\tau)=C(\epsilon_g^B,x;\tau)-C(\epsilon_g^A,x;\tau)$, where correlations are computed after smoothing with an averaging window of duration $\tau$. 
Positive values indicate that recalibration increases the coupling of gate error to parameter $x$ at timescale $\tau$, while negative values indicate that this coupling is stronger for the static baseline. 
The dashed vertical line at $\tau \approx 60\,\mathrm{s}$ marks the averaging window used for the correlations shown in Fig.~\ref{fig:3}(g), chosen to be the same as the averaging window used in Figs.~\ref{fig:3}(d,e,f). 
Across all $\tau$, recalibration consistently strengthens the correlation with coherence, $\Delta C(\Gamma_1;\tau)>0$ (green line), and its non-monotonic structure tracks the Allan-deviation feature of $\Gamma_1$ (orange line). 
By contrast, $\Delta C(\Delta f_d;\tau)<0$ over all $\tau$, showing that continuous recalibration suppresses the influence of qubit-frequency drift on gate error in a largely timescale-independent way. 
For $A_\pi$ and $A_{\pi/2}$, $\Delta C(x;\tau)$ is positive for short $\tau$ but becomes negative at longer $\tau$, consistent with the static baseline accumulating sensitivity as amplitudes drift away from their initial values, in this case on a timescale of $\tau \approx 10\,\textrm{s}$.

Taken together, these results show that our on-FPGA recalibration scheme tracks laboratory-time drift of performance parameters and continually correct those errors to optimize quantum processor performance over orders-of-magnitude timescales.

%%%%%%%%%%%%%%%%%%%%%
%%    FIGURE 4     %%
%%%%%%%%%%%%%%%%%%%%%
\section{Timing and Round-trip Latency Limits} 
\label{sec:timing}

\begin{figure}[t]
    \centering
    \includegraphics[width=\linewidth]{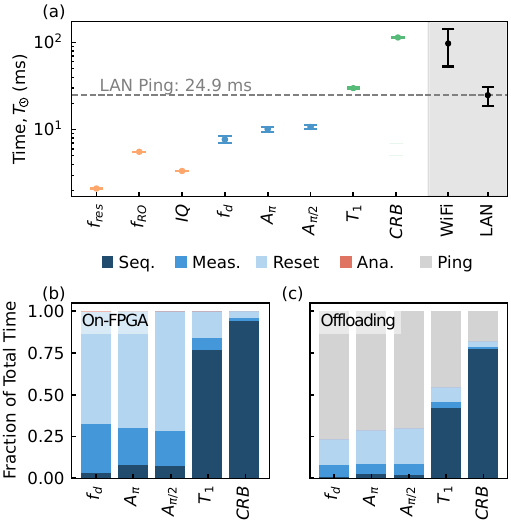}
\caption{Timing budget and latency limits. (a) Measured execution times $\Tstop$ for FPGA routines compared with representative WiFi/LAN round-trip delays,
highlighting the communication bottleneck for offloading-based analysis.
Here, $f_\mathrm{res}$ is a resonance-finding spectroscopy primitive using golden-section search, analogous to the $f_{01}$ peak-finding procedure in Fig.~2(d);
$f_\mathrm{RO}$ is the readout-frequency calibration shown in Fig.~2(a);
and IQ denotes a single IQ-discrimination update iteration, analogous to Fig.~2(c).
$T_1$ and CRB are extracted using the same ADE-style procedure.
Error bars indicate $1\sigma$ uncertainty estimated via bootstrapping over repeated measurements ($3000$ repeats for the experiment timings and $1000$ repeats for the network round-trip measurements).
(b) Timing breakdown example for some of the on-FPGA calibration primitives, showing that the wall-time is dominated by qubit reset and acquisition rather than classical analysis (relative time not visible at this scale).
(c) Timing breakdown example of representative instantiation of the traditional offloading procedure, which is dominated by network communication.
}

\label{fig:4}
\end{figure}

By relocating experimental analysis to the on-FPGA workflow and sampling more sparsely, we achieve a substantial reduction in the time-to-decision $\Tstop$. 
Once this improvement is realized, the limiting factors shift from classical overheads to experiment-level contributions. 
To identify the dominant contributors to calibration speed, we profile the timing budget of each primitive, as shown in Fig.~\ref{fig:4}(a)--(c). 

Figure~\ref{fig:4}(a) compares $\Tstop$ across a range of on-FPGA primitives with the measured round-trip latency over WiFi and LAN.
We define the round-trip latency as the delay incurred when measurement outcomes are transferred from the controller to a separate computer (typically, a measurement computer) and a response is returned to the controller, i.e., the communication overhead imposed by the offloading workflow.
We measure average round-trip latencies of $105\,\mathrm{ms}$ (WiFi connection between controller and measurement computer) and $25\,\mathrm{ms}$ (LAN connection).
In Fig.~\ref{fig:4}(a), $f_\mathrm{res}$ is determined via a golden-section search peak-finding routine (analogous to the $f_{01}$ peak-finding in Fig.~2(d)),
$f_\mathrm{RO}$ follows the readout-frequency calibration procedure in Fig.~2(a),
IQ denotes a single IQ-discrimination update iteration (Fig.~2(c)),
and $T_1$ and CRB are extracted using the ADE-style workflow.
Error bars show $1\sigma$ bootstrapped uncertainty from repeated measurements ($3000$ repeats for experiment timings; $1000$ repeats for network round-trip measurements).
All on-FPGA primitives complete below the LAN round-trip limit, with the exception of $T_1$ and CRB.

Figure~\ref{fig:4}(b) decompose $\Tstop$ into contributions from pulse-sequence execution (Seq.), measurement time (Meas.), qubit reset (Reset), analysis time (Ana.), and communication latency (Ping) as fractions of the total execution time-to-decision of each primitive in the on-FPGA workflow. 
Analysis constitutes a negligible contribution of only a few microseconds on average ($\sim 1\,\mu\mathrm{s}$). 
Calibration primitives of pulse frequency and amplitudes are dominated by qubit-reset latency, due to a repeat-until-success fast-feedback loop, which may require several sequential measurements to complete. 
While we use such active qubit reset in this work, the same primitives are compatible with restless operation ~\cite{Rol2017RestlessGates,Werninghaus2021High-SpeedReset,Tornow2022MinimumRates}, which would further increase repetition rate and calibration throughput. 
Meanwhile, the characterization primitives of $T_1$ and CRB are dominated by longer sequence lengths, on the order of the coherence time of the qubit, required to reach accurate benchmarking statistics.

In Fig.~\ref{fig:4}(c) we execute identical calibration primitives in the offloading workflow with sparse sampling using ADE and SPE, but performing the analysis on a separate measurement computer, connected via LAN.
Comparing Fig.~\ref{fig:4}(b) and Fig.~\ref{fig:4}(c), it is clear that although the FPGA \textit{analysis} itself contributes only a few microseconds on average ($\sim 1\,\mu\mathrm{s}$), the act of offloading itself becomes the ultimate time bottleneck when not deploying on-FPGA workflows.

%%%%%%%%%%%%%%%%%%%%%
%%    FIGURE 5     %%
%%%%%%%%%%%%%%%%%%%%%

\section{Uncertainty Scaling Under Sparse Sampling}
\label{sec:uncertainty}
\begin{figure}[t]
    \centering
    \includegraphics[width=\linewidth]{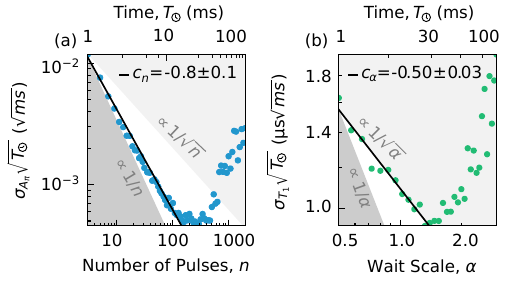}
    \caption{Uncertainty scaling under sparse sampling. (a) Analysis of the $\pi$-train amplitude-correction estimator as a function of total measurement time (or equivalently the number of pulses). 
    Uncertainty is estimated by bootstrapping and decreases with increased number of pulses, $n$, following a $1/n$ Heisenberg-like scaling.
    At high $n$, the analysis breaks down because the length of $\pi$-pulses approaches $T_1$.
    (b) Scaling for sparse $T_1$ estimation as a function of wait time scale factor ($\alpha$). 
    For small-to-moderate $\alpha$, the statistical uncertainty (from bootstrapping) follows a $1/\sqrt{\alpha}$ trend, but exhibit systematic deviations for long wait time scale factors (see text for details).
    }
    \label{fig:5}
\end{figure}

Beyond wall-time latency, a calibration loop must reach sufficient estimation precision within the available time-to-decision budget. 
Experimental execution time scales linearly with the number of measurement shots and the sequence duration.
Hence, the practical question is how to choose experimental settings to minimize estimation uncertainty for a target calibration time-window. 
In Fig.~\ref{fig:5} we quantify this for two representative primitives, namely pulse-amplitude correction using SPE with a $\pi$-train estimator (Fig.~\ref{fig:5}(a)) and $T_1$ estimation using ADE (Fig.~\ref{fig:5}(b)).

Throughout this section, uncertainties are bootstrapped from the measurements shots using 300 bootstrap replicates~\cite{Efron1979BootstrapJackknife}, and we report the time-normalized uncertainty $\sigma\sqrt{\Tstop}$ to compare different time-to-decision settings on equal footing.
We use bootstrapping here (rather than purely analytical error propagation) to capture non-idealities and estimator breakdown when the underlying signal model or its assumptions fail, e.g., at long $\pi$-trains or large $\alpha$ where coherence and finite-visibility effects become important.
In contrast, in Secs.~\ref{sec:drift}-\ref{sec:toolkit} we focus on regimes where the analytical error-propagation expressions are valid and use them because they are fast and suitable for real-time uncertainty estimates on the controller.

We first analyse the $\pi$-train amplitude-correction primitive in Fig.~\ref{fig:5}(a). 
We use SPE (Section~\ref{sec:3pointsampling}) for successively longer sequences of $n$ $\pi$-pulses.
We fit the (scaled) uncertainty to a general power law of the form $\propto n^{c_n}$, and find a fitted exponent $c_n = -0.8 \pm 0.1$, corresponding to an uncertainty scaling of roughly $1/n$. 
This scaling is consistent with error-amplifying $\pi$-train protocols, where increasing $n$ increases sensitivity to amplitude errors and therefore the information gained per unit time.
Ultimately, the scaling breaks down (around $n\approx 150$), due to a combination of coherence limiting the signal visibility (the pulse train for $n$ is approximately 6~$\mu$s long) and finite sampling at the amplitude axis leading to aliased response function of the $\pi$-train.

We next apply the same procedure to the sparse $T_1$ primitive in Fig.~\ref{fig:5}(b). 
Here, we use ADE to estimate $T_1$ with $t_0=16$ns and $\Delta t=\alpha \tilde{T}_1$, where $\alpha$ is a dimensionless wait scale and $\tilde{T}_1$ is our latest estimate of $T_1$ as in Section~\ref{sec:ADE}.
For any fixed $\alpha$, the estimator is shot-noise limited and we therefore expect the usual $\sigma \propto 1/\sqrt{\Tstop}$ scaling; however, varying $\alpha$ changes the information gained per unit time (i.e., the prefactor in this scaling).
We again fit a general power law to the (scaled) uncertainties, and find a fit parameter $c_\alpha = -0.50 \pm 0.03$, corresponding to an uncertainty scaling roughly $\propto 1/\sqrt{\alpha}$.
At larger wait times ($\alpha\gtrsim 2$), the normalized uncertainty increases, since the ADE points at these wait times fall deep in the tail of the exponential decay, reducing the information gain about the decay rate.

%%%%%%%%%%%%%%%%%%%%%
%%   CONCLUSION    %%
%%%%%%%%%%%%%%%%%%%%%

\section{Conclusion}
We have demonstrated an on-FPGA workflow in which pulse generation, measurement, analysis, and parameter feed-forward are co-located on control hardware to minimize time-to-decision, avoiding the communication overhead of the offloading workflow. 
A central result enabling our work is the introduction of two analytical tools amenable to memory-efficient FPGA implementation. 
Analytical decay estimation (ADE) provides a closed-form three-point estimator for exponential responses, enabling rapid coherence inference and reuse for both $T_1$ estimation and randomized benchmarking. 
Complementarily, sparse phase estimation (SPE) yields an analytic update rule for sine-like response functions, providing a robust, low-overhead primitive for fast pulse-amplitude tracking. 
Deployed in closed loop over six hours with in situ CRB validation, frequent recalibration improves benchmarking performance by a $6.4\%$ reduction in gate infidelity relative to a static baseline while tracking concurrent drift in coherence and control parameters. 
Timescale-resolved correlation analysis shows that continuous recalibration largely removes the influence of detuning and amplitude drift on gate error, leaving a gate-performance that is tightly tracking instantaneous coherence properties of the device.
Finally, timing and uncertainty profiling clarifies the practical limits of millisecond calibration. 
In our implementation, the execution time of calibration experiments were limited by our active reset scheme while characterization experiments were limited by sequence duration.
For short durations, uncertainty follows the averaging scalings, while at longer durations non-stationary drift and accumulated imperfections introduce systematic deviations that set an effective floor. 
Together, these results motivate extending on-FPGA analysis to additional primitives and multi-qubit routines. 
Our work thus establishes a practical route to calibration schedules that balance time-to-decision against estimator uncertainty and support autonomous millisecond-timescale operation in drifting superconducting qubit devices.

%%%%%%%%%%%%%%%%%%%%%
%% ACKNOWLEDGMENTS %%
%%%%%%%%%%%%%%%%%%%%%
\begin{acknowledgments}
We thank Lars Lemming, Subhadip Das and the electronics and machine shops at the University of Copenhagen for support with the laboratory establishment.

This research was supported by the Novo Nordisk Foundation (grant no. NNF22SA0081175), the NNF Quantum Computing Programme (NQCP), Villum Foundation through a Villum Young Investigator grant (grant no. 37467), the Innovation Fund Denmark (grant no. 2081-00013B, DanQ), the U.S. Army Research Office (grant no. W911NF-22-1-0042, NHyDTech), by the European Union through an ERC Starting Grant, (grant no. 101077479, NovADePro), by the Carlsberg Foundation (grant no. CF21-0343), HORIZON-MSCA-2024-PF-01 (grant no.101204890, QuARTEC).
Any opinions, findings, conclusions or recommendations expressed in this material are those of the author(s) and do not necessarily reflect the views of Army Research Office or the US Government. 
Views and opinions expressed are those of the author(s) only and do not necessarily reflect those of the European Union or the European Research Council. Neither the European Union nor the granting authority can be held responsible for them. 
Finally, we gratefully acknowledge Lena Jacobsen and Helle Grunnet for program management support.

\end{acknowledgments}

%%%%%%%%%%%%%%%%%%%%%
%%    REFRENCES    %%
%%%%%%%%%%%%%%%%%%%%%
\bibliography{references}

%%%%%%%%%%%%%%%%%%%%%
%%    APPENDIX     %%
%%%%%%%%%%%%%%%%%%%%
\onecolumngrid
\newpage
\appendix
% Appendix / Supplementary Methods

\setcounter{figure}{0}
\renewcommand{\thefigure}{S\arabic{figure}} 
\setcounter{table}{0}
\renewcommand{\thetable}{S\arabic{table}}
\setcounter{equation}{0}
\renewcommand{\theequation}{S\arabic{equation}}

\section{Experimental apparatus and fabrication}\label{app:experimental-setup}
\begin{figure}[H]
    \centering
    \includegraphics[width=\linewidth]{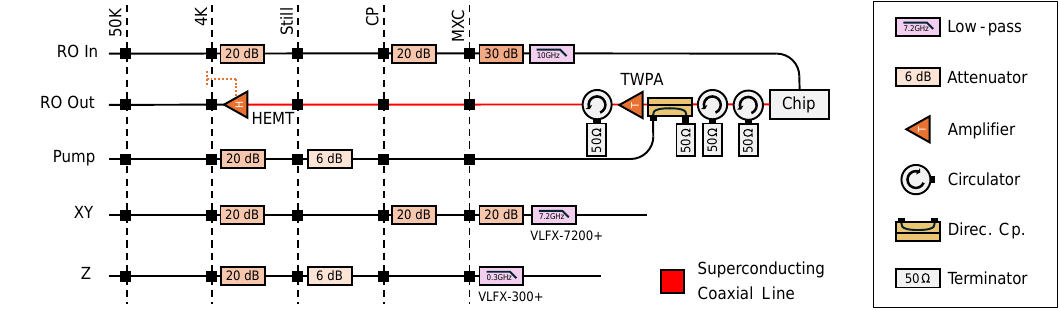}
\caption{Wiring schematic of our experimental setup. The fridge wiring is optimized according to the results of~\cite{Krinner2019EngineeringSystems}.}
\label{fig:fridge_diagram}
\end{figure}
Our experimental setup consists of a 12-qubit superconducting quantum processor (see Fig.~\ref{fig:1}(c)) packaged inside of a QCage64 sample holder from Quantum Machines. 
The package is mounted to the mixing chamber of a Bluefors XLD1000 dilution refrigerator. 
The qubits are flux-tunable transmons with fixed direct qubit-qubit capacitive coupling. 
The wiring of the fridge is shown in Fig.~\ref{fig:fridge_diagram}. 
Our control hardware consists of a Quantum Machines OPX1000 chassis containing front-end modules (FEMs) for controlling both the XY-control and readout microwave signals (MW-FEM) and low frequency (LF-FEM) for baseband flux control for the Z-control lines. 
The TWPA pump is provided by a Rohde \& Schwarz SGS100a.
The Quantum Machines hardware gives access to some parts of the FPGA through the QUA intermediate language API. 
Using the mathematical tools provided by the QUA language, we extend the API by implementing the toolbox directly on the FPGA, including optimizers, on-board analysis, and feedback routines. 
All experiments are orchestrated with our in-house python data acquisition platform called \textit{Pelagic}.

The device itself is fabricated on a four-inch, high-resistivity intrinsic silicon wafer.
The substrate underwent a standard cleaning process using Piranha solution followed by a Hydrofluoric acid (HF) dip to remove native oxides.
A 200\,nm Aluminum (Al) base layer was deposited via e-beam evaporation (Plassys MEB550S).
The ground plane, including control circuitry and large qubit structures, is patterned with optical lithography and subsequently wet-etched in Transene Type D etchant at $50^{\circ}$\,C.
The wafer is then diced into 2.2\,cm\,$\times$\,2.2\,cm chips, after which we deposit Manhattan-style Josephson junctions using double-angle shadow evaporation with a bilayer resist stack. 
This is performed in the same Plassys tool that the ground plane was deposited in. 
We evaporate 40\,nm/80\,nm Al for the first/second junction electrode with an intermediate in-situ static oxidation step for 6 minutes at 120 mbar.
We deposit a 200\,nm Al patch after an in-situ Argon ion-milling step to ensure electrical connectivity of the junctions to the qubit capacitive pads.
We finally deposit airbridges to the device to eliminate spurious slotline modes in the coplanar waveguides which carry signals to the device, to mitigate crosstalk between various components, and ensure a well-defined ground across the device.
We spin AZ5214E and pattern/develop the airbridge anchor points.
The airbridge scaffolding is formed by reflowing the resist at $140^{\circ}$\,C. 
We then deposit a layer of 400\,nm Al in the Plassys after ion-milling the anchor points. 
The bridge is then defined by performing a second round of optical lithography and the aluminum is wet-etched. 
The optical resist is then stripped and the device is cleaned via oxygen plasma ashing.

\section{Downsampling and moving average} \label{app:downsampling}
\begin{figure}[H]
    \centering
    \includegraphics[width=.5\linewidth]{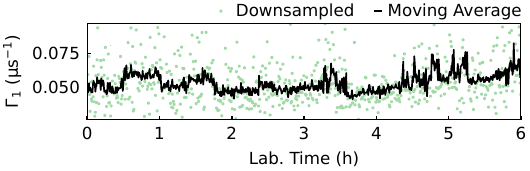}
\caption{Comparing moving average and downsampling of dense data. Moving average smoothing (black) uses a sliding window 100 points to suppress high-frequency fluctuations while preserving the original point density. 
Downsampling (green points) retains one estimate every 100 updates, illustrating the series obtained at a lower effective update rate.}
\label{fig:gamma1_timeseries}
\end{figure}
In Fig.~\ref{fig:gamma1_timeseries} we distinguish between \emph{moving (rolling) average smoothing} and \emph{downsampling}, which address different questions when visualizing the dense data presented in Figure~\ref{fig:3} of the main paper. 
For a moving average, we replace each point by the average over a sliding window of experimentally recorded estimates.
In the case of Fig.~\ref{fig:gamma1_timeseries} we choose a rolling window of 100 sequential datasets (black points).
In this situation, adjacent samples share most of the same underlying shots and the resulting curve is strongly correlated in time.
This operation primarily suppresses high-frequency fluctuations while preserving the original sampling density and is therefore well suited for revealing slow trends without changing when estimates are produced. 
By contrast, downsampling retains only every $N$th estimate, reducing the number of displayed points and yielding samples that are far less correlated because they are built from non-overlapping data.
For Fig.~\ref{fig:gamma1_timeseries} we show the same underlying dataset, downsampled by a factor 100.
Downsampling therefore represents what the time series would look like if the calibration loop were actually executed at a lower update rate (in this case, $\times $100 slower), whereas a moving average is a visualization choice that smooths the same underlying high-rate stream. 
The black points in Fig.~\ref{fig:gamma1_timeseries} have first been smoothed via moving average, and the resulting data has then been downsampled by a factor of $100$, and the green points represent only downsampling without smoothing.

\section{Readout SNR and IQ discrimination}
\label{app:readout-snr}
Here we detail the definition of SNR in the context of IQ, used as the scalar objective for the on-FPGA IQ discriminator.
At each candidate readout setting (i.e., for a given choice of readout detuning and amplitude), we acquire batches of single-shot, integrated IQ samples, \(\{(I,Q)\}\), for two state-preparation experiments, $\lvert 0\rangle$ and $\lvert 1\rangle$.
% These IQ samples define the scalar objective optimised in the readout routine and they provide the training data for the IQ discriminator.
For readout optimisation we quantify how well separated the two IQ clouds are using a scalar signal-to-noise ratio (SNR),
\begin{equation}
\mathrm{SNR} \equiv
\frac{\left\lVert \boldsymbol{\mu}_1-\boldsymbol{\mu}_0 \right\rVert}{\sqrt{\sigma_0^2+\sigma_1^2}} .
\end{equation}
Here, \(\boldsymbol{\mu}_k=(\overline{I}_k,\overline{Q}_k)\) is the centroid of the $\lvert k\rangle$ IQ cluster (computed from the sample means), and \(\sigma_k^2\) is the corresponding radial variance of the single-shot, integrated IQ samples.
The Nelder--Mead optimiser treats this SNR as its objective and iteratively proposes new readout parameters to maximise it.

In addition, at each optimisation step we use a simple $n$-state IQ classifier (here $n=2$) to label individual shots.
Given a measurement vector $\mathbf{x}=(I,Q)$, we form a variance-normalised squared distance to each class $k$,
\begin{equation}
d_k^2(\mathbf{x}) =
\frac{(I-\overline{I}_k)^2+(Q-\overline{Q}_k)^2}{\sigma_k^2} ,
\end{equation}
where \((\overline{I}_k,\overline{Q}_k)\equiv \langle (I,Q)\rangle_k\) and
\(\sigma_k^2 \equiv \left\langle (I-\overline{I}_k)^2+(Q-\overline{Q}_k)^2\right\rangle_k\)
are estimated from the corresponding training batches.
We then assign each shot to a $|0\rangle$ or $|1\rangle$ outcome based on the smallest variance-normalised distance,
\begin{equation}
\hat{k} = \underset{j\in\{0,\ldots,n-1\}}{\arg\min}\,d_j^2(\mathbf{x}).
\end{equation}

\section{Optimiser stopping criteria and iteration counts}
\label{app:stopping}
In the main text we report typical convergence in \(\sim 20\) Nelder--Mead iterations.
In practice we do \emph{not} enforce a fixed iteration cap; instead, we terminate when the simplex contracts below a
preset tolerance and the objective improvement falls below a preset threshold (both implemented directly on the FPGA).
This makes the reported iteration counts an empirical convergence property under our measurement conditions rather than a hard limit.

\section{Golden-section search (GSS) summary}
\label{app:gss}
Here we briefly review the algorithm of Golden-section search.
Golden-section search is a bracketing method for one-dimensional optimisation of a unimodal objective function.
Starting from an interval \([a,b]\) containing the optimum, GSS evaluates the objective at two interior points whose
spacing is set by the golden ratio, then discards the sub-interval that cannot contain the optimum.
Repeating this procedure shrinks the bracket exponentially while requiring only one new function evaluation per
iteration.
This makes GSS well suited for FPGA execution, where we wish to minimise both control-flow complexity and the total
number of measurements. 

\section{Analytical Decay Estimator}
\label{app:sparse-exp}
In this section we derive the form of the analytical 3-point estimator for exponential decays (Analytical Decay Estimator, ADE).
In the main text, ADE is used to track the decay of exponential signals without having to calibrate possible offsets or amplitudes of the signal coming from SPAM. 
In general, exponential signals are of the form: 
\begin{equation}
P_{\ket{1}}(t) = Ae^{-\Gamma_1 t} + C,
\end{equation}
and in most cases we are exclusively interested in the decay parameter $\Gamma_1$.
We estimate the decay rate by constructing an estimator that cancels all other unwanted parameters, while still being analytically solvable for $\Gamma_1$ from measurements of the signal. 
We measure the signal $ P_{\ket{1}}(\tau)$ at delays $\tau \in \{t_0,\, t_0+\Delta t,\, t_0+3\Delta t\}$ and consider the fraction $c(t_0, \Delta t)$: 
\begin{equation}\label{eq:Third_order_eq}
c(t_0, \Delta t) \equiv \frac{P_{\ket{1}}(t_0 + 3\Delta t) - P_{\ket{1}}(t_0)}{P_{\ket{1}}(t_0 + \Delta t) - P_{\ket{1}}(t_0)} = \frac{Ae^{-\Gamma_1(t_0 + 3\Delta t)} + C - \left(Ae^{-\Gamma_1 t_0} + C\right)}{Ae^{-\Gamma_1(t_0 + \Delta t)} + C - \left(Ae^{-\Gamma_1 t_0} + C\right)} = \frac{x^3 - 1}{x - 1} = x^2 + x + 1,
\end{equation}
where $x\equiv e^{-\Gamma_1\Delta t}$. For known $c(t_0,\Delta t)$, Eq.~\ref{eq:Third_order_eq} gives a second-order polynomial equation in $x$ with solution 
\begin{equation}
    x = \sqrt{c(t_0, \Delta t)-\frac{3}{4}} - \frac{1}{2}.
\end{equation}
Which can be verified by explicit insertion. From this and the definition of $x$, we find the rate to be 
\begin{equation}
\Gamma_1 =
-\frac{1}{\Delta t}\ln\!\left(\sqrt{c(\Delta t, t_0) - \frac{3}{4}} - \frac{1}{2}\right),
\end{equation}
with \(T_1 = 1/\Gamma_1\). 

Notice that Eq.~\ref{eq:Third_order_eq} is only analytically solvable in $x$ for certain choices of delay times $\tau$. Here, we choose delays such that the equation became quadratic in $x$ and analytically solvable. This choice was made under speed considerations: There is more information about $\Gamma_1$ in the tail of the exponential than in the beginning, so choosing the last delay time to be larger gives us better confidence of the estimate of $\Gamma_1$.

Had we chosen $\tau_2 = t_0 + n\Delta t$, we would have gotten an $(n-1)$-th order polynomial equation in $x$ that is analytically solvable for $n\leq 6$, but with solutions that are increasingly convoluted. 
Balancing analytical simplicity for information gain, we opted for the solution with $n=3$. 

In the main text we use this estimator to infer coherence times and gate fidelities, but the same construction applies generally to extracting the characteristic decay constant (or decay time) of any observable that follows an exponential dependence on a control parameter (e.g., time, sequence length, or circuit depth). 

\section{Sparse Phase Estimator}\label{app:sparse-sinus}
In this section, we derive the form of the Sparse Phase Estimator used in the main text. We construct an analytical 3 point estimator to track  phase corrections of sinusoidal signals without having to calibrate possible offsets or amplitudes of the signal coming from SPAM. Such signals will be of the form: 
\begin{equation}
  P_{\ket{1}}(\theta) \;=\; A\cos\!\big(\theta) + C,
\end{equation}
Where $A$ and $C$ are SPAM-parameters and $\theta$ is an angle-parameter of the signal. In the SPE-framework, we assume to have some primitive with angle-parameter $\theta_0$ that we wish to determine. In order to do so, SPE assumes that we have sample-access to the signal at $\theta_0$ and at $\theta_\pm = \theta_0 \pm \pi/2$.
Define the average $\overline{P}_{\ket{1}}=\tfrac{1}{2}\!\left(P_{\ket{1}}(\theta_-)+P_{\ket{1}}(\theta_+)\right)$. In order to eliminate SPAM-parameters, we construct the fraction
\begin{equation}
  R \;\equiv\;
  \frac{P_{\ket{1}}(\theta_-) - P_{\ket{1}}(\theta_+)}
       {\,2\left(P_{\ket{1}}(\theta_0) - \overline{P}_{\ket{1}}\right)}.
\end{equation}
Using angle-sum identities,
\[
\begin{aligned}
  P_{\ket{1}}(\theta_-) - P_{\ket{1}}(\theta_+)
  &= 2A\,\sin(\theta_0)\\
  P_{\ket{1}}(\theta_0) - \overline{P}_{\ket{1}}
  &= A\,\cos(\theta_0)\,
\end{aligned}
\]
so the ratio simplifies to
\begin{equation}
  R \;=\; \tan(\theta_0)
\end{equation}
This suggests a phase estimator of the signal
\begin{equation}
  \widehat{\theta} \;=\; \operatorname{atan2}\!\Big(P_{\ket{1}}(\theta_-) - P_{\ket{1}}(\theta_+),\;
  2\big(P_{\ket{1}}(\theta_0) - \overline{P}_{\ket{1}}\big)\Big)
\end{equation}
This estimator is the one we've been using in most examples in the main text. It then directly gives an estimate of the correction to the angle $\widehat{\delta\theta} = \widehat{\theta} - \theta_{\text{targ}}$, if we have some target angle $\theta_{\text{targ}}$. 

As an illustrative example, we can use SPE to estimate the detuning in a Ramsey-Chevron experiment (as done in Fig.\ref{fig:3}(e) of the main paper). 
In the case of Ramsey-Chevron, the signal will be of the form
\begin{align}
    S(f) = A\cos(2\pi\Delta f \tau) + C
\end{align}
where $\Delta f$ denotes a frequency detuning from the 0-1 transition and $\tau$ denotes the Ramsey delay duration. So by sampling at detuning corresponding to the prior best-estimate for the frequency
$\Delta f_0$ and the two off-set detunings $\Delta f_\pm = \Delta f_0 \pm \frac{1}{4\tau}$, this corresponds to measuring at the SPE-prescribed points of measurement $\theta_0$ and $\theta_\pm = 2\pi \Delta f_\pm \tau = \theta_0 \pm \pi/2$ for the full argument $\theta \equiv 2\pi\Delta f \tau$ of the signal. 
Hence, by learning $\theta_0$ in SPE, we can reconstruct the detuning of the signal,
\begin{align}
\Delta f_0 = \frac{\theta_0}{2\pi\tau}.
\end{align}

In instances, such as $\pi$-pulse amplitude-calibration, it is no longer true, that we can sample the angle variable at $\theta_\pm = \theta_0 + \pi/2$. To understand why, we briefly examine the equation of the signal, 
\begin{align}
    S(\alpha) = A\cos(n_\pi \alpha) + C = A\cos(n_\pi \Omega t) + C
\end{align}
where $\alpha = \Omega t$ is the rotation angle on the bloch-sphere from a single application of a $\pi$-pulse, $t$ is the signal duration and $\Omega$ is the Rabi frequency of the oscillation. In experiment, we have access to the drive amplitude of the signal $V_0$, but as the rabi frequency is only linearly related to this property $\Omega = a\cdot V_0$ for unknown proportionality constant $a$, we see that we cannot sample the signal at the SPE-prescrived arguments ($\theta_\pm = n_\pi \alpha \pm \pi/2$).
We must therefore vary the analysis slightly and assume a different access model. 

We propose to then instead sample at the modified points $\alpha_0$ and $\alpha_{\pm} =\alpha_0\left(1\pm\frac{1}{2n_\pi}\right)$, where $\alpha_0$ is the rotation angle from our $\pi$-pulse primitive. This will approximate the same analysis as above, given that $\alpha_0\approx \pi$. 
The resulting correction rule is written in Eq.~\eqref{eq:pi-correction-3PE} in the main text.

The same analysis works when estimating the amplitude of a $\pi/2$-pulse by applying the pulse twice and dividing the estimated angle by two.

\end{document}